\documentclass[superscriptaddress,prl,twocolumn,amsmath,amssymb]{revtex4}
\usepackage{graphicx}
\usepackage{dcolumn}
\usepackage{bm}

\usepackage{amsmath}

\newcommand{\ket}[1]{\left | #1 \right \rangle}

\begin{document}

\title{High-fidelity operation of quantum photonic circuits\vspace{-8pt}}

\author{Anthony Laing}
\affiliation{Centre for Quantum Photonics, H. H. Wills Physics Laboratory \& Department of Electrical and Electronic Engineering, University of Bristol, Merchant Venturers Building, Woodland Road, Bristol, BS8 1UB, UK}

\author{Alberto Peruzzo}
\affiliation{Centre for Quantum Photonics, H. H. Wills Physics Laboratory \& Department of Electrical and Electronic Engineering, University of Bristol, Merchant Venturers Building, Woodland Road, Bristol, BS8 1UB, UK}

\author{Alberto Politi}
\affiliation{Centre for Quantum Photonics, H. H. Wills Physics Laboratory \& Department of Electrical and Electronic Engineering, University of Bristol, Merchant Venturers Building, Woodland Road, Bristol, BS8 1UB, UK}

\author{Maria Rodas Verde}
\affiliation{Centre for Quantum Photonics, H. H. Wills Physics Laboratory \& Department of Electrical and Electronic Engineering, University of Bristol, Merchant Venturers Building, Woodland Road, Bristol, BS8 1UB, UK}

\author{Matthaeus~Halder}
\affiliation{Centre for Quantum Photonics, H. H. Wills Physics Laboratory \& Department of Electrical and Electronic Engineering, University of Bristol, Merchant Venturers Building, Woodland Road, Bristol, BS8 1UB, UK}

\author{Timothy C. Ralph}
\affiliation{Department of Physics, University of Queensland, St Lucia, Queensland 4072, Australia.}

\author{Mark G. Thompson}
\affiliation{Centre for Quantum Photonics, H. H. Wills Physics Laboratory \& Department of Electrical and Electronic Engineering, University of Bristol, Merchant Venturers Building, Woodland Road, Bristol, BS8 1UB, UK}

\author{Jeremy L. O'Brien}\email{Jeremy.OBrien@bristol.ac.uk}
\affiliation{Centre for Quantum Photonics, H. H. Wills Physics Laboratory \& Department of Electrical and Electronic Engineering, University of Bristol, Merchant Venturers Building, Woodland Road, Bristol, BS8 1UB, UK}

\begin{abstract}We demonstrate photonic quantum circuits that operate at the stringent levels that will be required for future quantum information science and technology. These circuits are fabricated from silica-on-silicon waveguides forming directional couplers and interferometers. While our focus is on the operation of quantum circuits, to test this operation required construction of a spectrally tuned photon source to produce near-identical pairs of photons. We show non-classical interference with two photons and a two-photon entangling logic gate that operate with near-unit fidelity. These results are a significant step towards large-scale operation of photonic quantum circuits.\end{abstract}
\maketitle

Quantum information science \cite{nielsen} is not only a fundamental scientific endeavor but 
promises profound new technologies in communication \cite{gi-rmp-74-145,gi-nphot-1-165}, information processing \cite{de-prsla-400-97,la-nat-464-45}, and ultra-precise measurement \cite{gi-sci-306-1330}.  However, as with their classical counterparts, these quantum technologies must be robust to imperfections in their components and to the effects of environmental noise.  For example, in the case of universal quantum computing, current estimates \cite{kn-nat-434-39,da-prl-96-020501,ra-prl-98-190504,da-pra-73-052306} of the maximum error rate per gate (EPG) range from a few \% to $10^{-4}$. Meeting these rigorous EPG requirements is a major challenge, owing to the fragility of quantum systems, and has thus far only been achieved in ion traps  \cite{be-natphys-4-463}.

Encoding quantum information in photons is promising for fast transmission, low intrinsic noise (or decoherence) and ease of implementing one-photon operations \cite{ob-nphot-3-687}. Consequently photons are the information carrier of choice for quantum communication \cite{gi-rmp-74-145,gi-nphot-1-165}. Realising the two-photon interactions required for the majority of quantum information protocols is more challenging, however, they can be achieved using only single photon sources, detectors and linear optical circuits \cite{kn-nat-409-46}, and much progress towards this goal has been made \cite{ob-sci-318-1567}. 
Integrated photonics---waveguide circuits lithographically patterned on-chip---holds great promise for miniaturizing and scaling quantum logic circuits \cite{po-sci-320-646,po-sci-325-1221,ma-oe-17-12546}, and high fidelity single-qubit operations have already been demonstrated \cite{matthews-2008}. However, the crucial two-qubit operations required for more general quantum information protocols have yet to be demonstrated at the required high fidelity levels.

\begin{figure}[tb]
  \includegraphics[width=0.8\columnwidth]{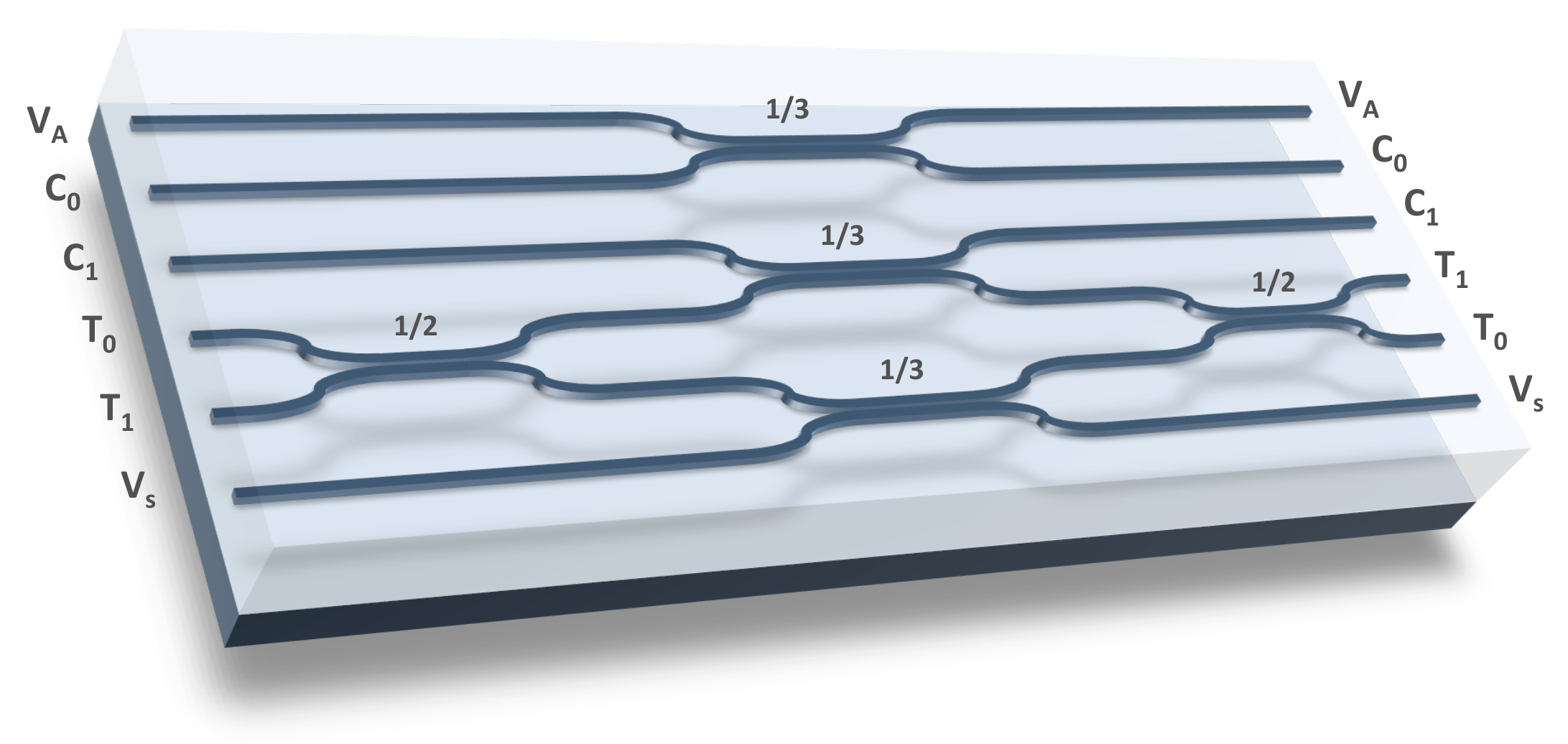}\\
  \caption{A silica-on-silicon waveguide quantum circuit. Directional couplers evanescently couple two waveguides---equivalent to a beamsplitter in bulk optics---where the length and separation of the waveguides in the coupling region determines the reflectivity $\eta$.  This CNOT gate is composed of three $\eta=1/3$ and two $\eta=1/2$ couplers to form a linear optical network with both classical and quantum interferometers, equivalent to bulk optical implementations \cite{ra-pra-65-062324,ho-pra-66-024308,ob-nat-426-264,ob-prl-93-080502}.
  }
  \label{wgs}
    \vspace{-0.5cm}
\end{figure}

Here we demonstrate integrated photonic devices that exhibit near-unit fidelity quantum interference and two-photon entangling logic operation: we observe a quantum interference or ``HOM" dip \cite{ho-prl-59-2044} with a minimum which reaches the ideal value, and a two-photon CNOT gate with a `logical basis fidelity' of $F=0.969\pm 0.002$ and similarity $S=0.993 \pm 0.002$, taking into account the deviation in the fabricated reflectivities of the directional couplers. 
Although our focus is on the operation of the circuits themselves, and not single photon sources or detectors,
observation of this high-fidelity operation relied on a spectrally tuned photon source producing near-identical pairs of photons. These results show that photonic quantum circuits can perform at the high fidelities required for future quantum technologies, and are likely to find application in fundamental scientific investigations where such high performance operation is required to observe uniquely quantum mechanical effects.

Quantum states are inherently fragile: typically, physical systems must be very small and very cold to exhibit the quantum phenomena of superposition and entanglement that lie at the heart of quantum information science and technology. Even in these extreme regimes, the state of a quantum system degrades due to unwanted interactions with its environment---decoherence---and imperfect operations on them---\emph{i.e.} initialisation, logic gates and measurement. This situation is exacerbated by the fact that quantum information is inherently analogue in nature, precluding the `latching' used in digital logic. 

Fortunately, errors can be encoded against by using quantum error correction \cite{shor96,st-prl-77-793}, whose complexity arises from the fact that directly measuring quantum systems disturbs them (which rules out naive majority error correcting codes for example) leading to the need for complicated entangled states of several particles to encode single logical states. 
The threshold theorem says that if the noise is below some threshold an arbitrarily long quantum computation can be realized \cite{kn-prsa-454-365}; any architecture that can work below this EPG threshold is said to be `fault tolerant', for the given error model. There are two broad classes of errors: locatable---essentially qubit erasure, caused by loss or gate failures; and unlocatable---bit flips \emph{etc}. Locatable errors are easier to fix and hence have a higher threshold (\emph{eg.} the  fault tolerance threshold considering only photon loss is at least 1/3 \cite{va-prl-100-060502}); here we address the more stringent thresholds corresponding to unlocatable errors. Even in cases where full error correction is not required, such as in quantum communication protocols, high-fidelity operation of fundamental building blocks is crucial to high performance operation of the given protocol.

\begin{figure}[tb]
\begin{center}
\vspace{-0.5cm}
  \includegraphics[width=0.6\columnwidth]{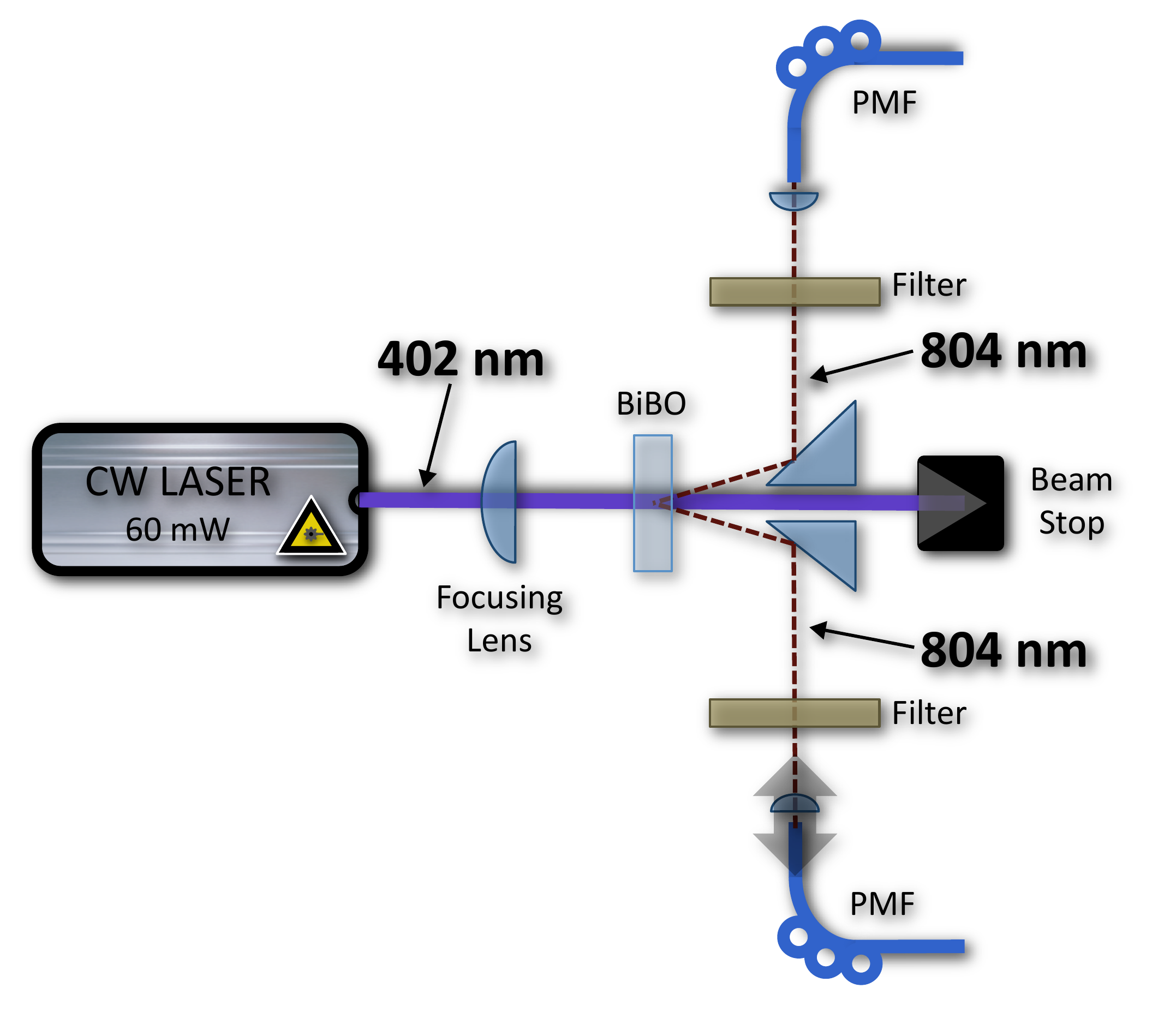}\\
  \caption{A schematic of the 
  two photon source. A CW laser was focused to a 40 $\mu$m waist in a Type-I BiBO crystal to produce degenerate photon pairs via spontaneous parametric down conversion.  The daughter photons were collected into polarisation maintaining fibres (PMFs) after passing through interference filters with a 2 nm FWHM.  The source was fine tuned with a spectrometer to ensure degenerate photons \cite{hifi-note-9}. 
  }
    \label{schematic}
  \end{center}
  \vspace{-0.7cm}
\end{figure}

In contrast to most systems---where fast coupling to the environment dominates---the major sources of error in photonic approaches to quantum information science and technology are photon loss, including source and detector inefficiency; unstable one-photon (`classical') interference, due to unstable phases (or path lengths) in optical circuits; and imperfect quantum interference, due to mode matching \cite{hifi-note-1}.
Progress towards high efficiency single photon sources \cite{sps,sh-nphot-1-215} and detectors \cite{spd,ha-nphot-3-696} is impressive; and integrated photonics holds great promise for miniaturizing and scaling high-performance photonic quantum circuits \cite{po-sci-320-646,po-sci-325-1221,ma-oe-17-12546,matthews-2008,sm-oe-17-13516}. 
While high fidelity single qubit operations have been demonstrated in this architecture  \cite{matthews-2008}, two photon logic gate operation, including quantum interference, below relevant EPG thresholds has not yet been demonstrated.

A silica-on-silicon waveguide circuit is shown in Fig.~\ref{wgs}. 
Photons are guided via total internal reflection due to a small refractive index contrast between the core of the waveguide and the surrounding cladding in the same way as a single mode optical fibre. Waveguides are brought within several $\mu$m to realize directional couplers whose reflectivity $\eta$ can be controlled via the waveguide separation or length of the coupling region (we use length). The major factors determining the performance of such a device are photon loss (typically $\sim$0.1 dB/cm); the quality of quantum interference \cite{ho-prl-59-2044} at directional couplers; and the quality of classical interference in interferometers formed by two or more directional couplers.

\begin{figure}[t]
\begin{center}
\vspace{-0.5cm}
  \includegraphics[width=0.7\columnwidth]{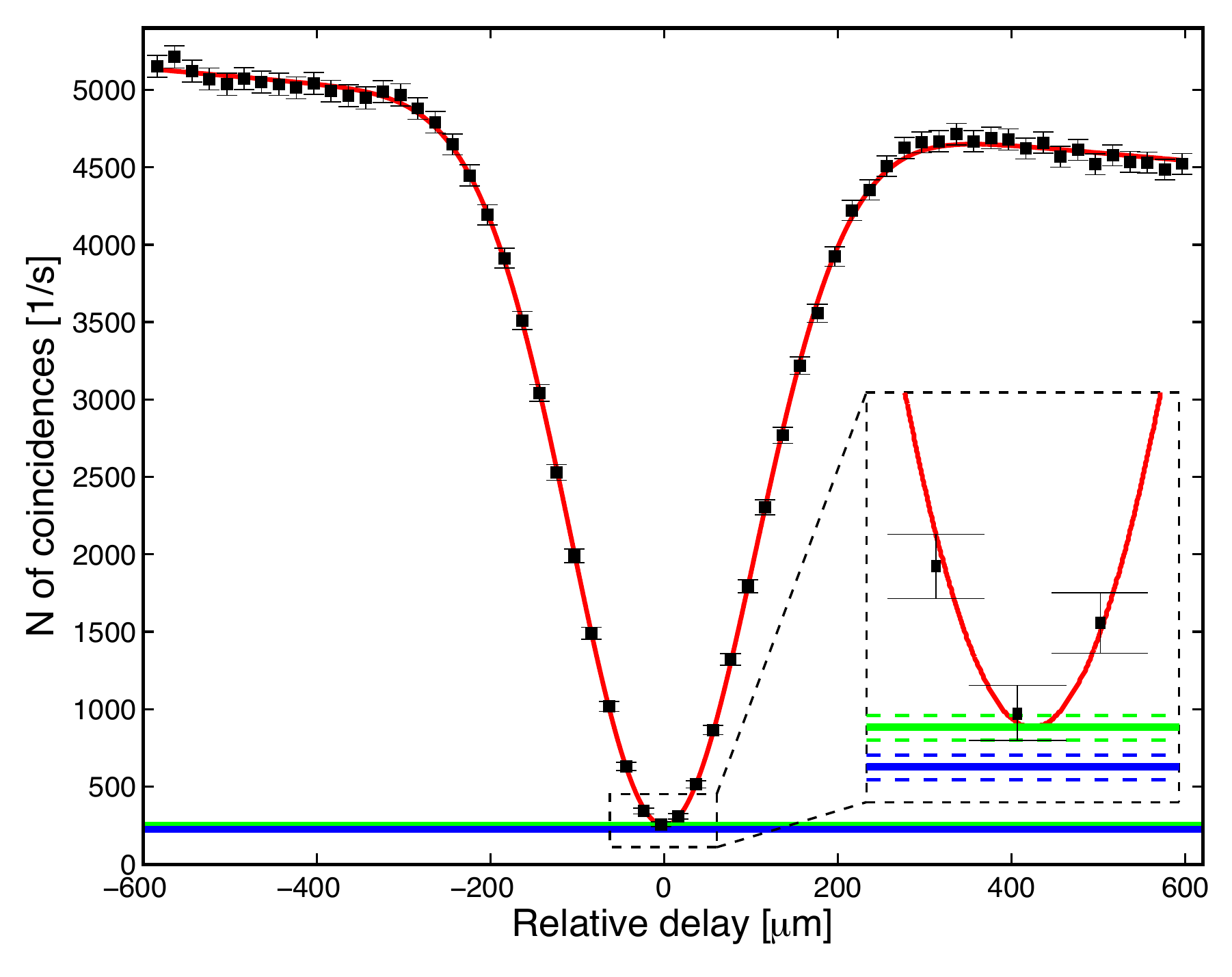}\\
  \caption{High fidelity quantum interference in a waveguide directional coupler.  The measured rate of detecting a photon at each output of a directional coupler is plotted as a function of the delay between the arrival of the photons at the coupler. 
The fit is a Gaussian with a linear term to account for the small decoupling as one arm of the source is translated to change the delay.  All error bars, arising from Poissonian counting statistics overlap the fit. The FWHM of 249.4 $\mu$m is as expected for the 2 nm interference filters used. The blue line shows the measured rate of accidental counts at the dip minimum position (with dashed error bars). The green line shows the count rate expected at the centre of the dip for the measured reflectivity $\eta=0.5267 \pm 0.0004$.}
    \label{hom}
  \end{center}
    \vspace{-0.5cm}
\end{figure}

\begin{figure*}[tb]
  \includegraphics[width=0.7\textwidth]{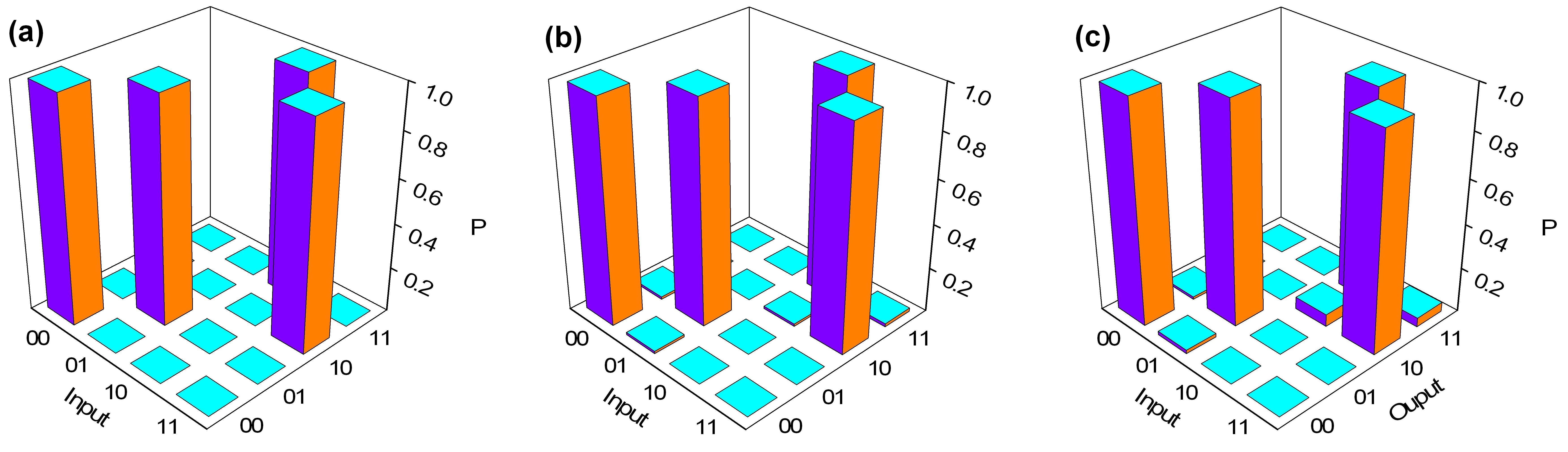}\\
  \caption{High-fidelity CNOT logic gate operation: (a)~The truth table of a CNOT gate. (b)~The ideal truth table for the measured device taking into account the measured $\eta$'s of the couplers, which differed slightly from the values given in Fig. \ref{wgs}(b). 
  (c) The experimentally measured truth table.
  }
\label{truth}
 \vspace{-0.3cm}
 \end{figure*}

Quantum interference \cite{ho-prl-59-2044} occurs when two photons simultaneously arrive at each input of a 
beamsplitter or directional coupler; for $\eta=0.5$ there is zero probability for one photon to be found in each output, since the photons exit in a superposition of both being in each output: $\ket{11}\rightarrow({\ket{20}-\ket{02}})/{\sqrt{2}}$.
This phenomenon arises due to destructive quantum interference of the two indistinguishable two-photon probability amplitudes---both photons reflected and both photons transmitted. Quantum interference lies at the heart of photonic quantum technologies: logic gates \cite{ra-pra-65-062324,ho-pra-66-024308,ob-nat-426-264,ob-prl-93-080502}, quantum filters \cite{sa-prl-96-083601, re-prl-98-203602, ok-sci-323-483}, Bell state analyzers \cite{la-prl-95-210504}, \emph{etc}. The degree of this interference is quantified by the visibility $V=({C_{class}-C_{quant}})/{C_{class}}$, 
where $C_{class}$ is the classical rate of detecting one photon in each output---experimentally measured by deliberately introducing a time delay such that the photons do not arrive simultaneously---and $C_{quant}$ is the experimentally measured rate for zero delay.

Quantum interference also occurs when $\eta\neq0.5$ with
\begin{equation}
\label{visibility}
V_{ideal}=\frac{2\eta(1-\eta)}{1 - 2 \eta + 2 \eta^2}.
\end{equation}
An experimentally measured visibility $V_{meas}<V_{ideal}$ arises due to any distinguishing information between the two two-photon amplitudes, including differences in the photons' polarization, spatial, spectral or temporal modes, or mixture in any degree of freedom.
Since $V_{meas}$ is limited by distinguishability it is critical that the photon source used to test a circuit produce photons that are highly indistinguishable.  Here we used the Type-I SPDC source shown schematically in Fig. \ref{schematic} \cite{hifi-note-9}. We emphasize that it is not our aim to test this SPDC source, but its construction was necessary to test the circuits.
 
We measured the rate of detecting a single photon at each output of an $\eta=0.5267 \pm 0.0004$ directional coupler as a function of the arrival time of the photons---plotted as black data, with the red fit, in Fig.~\ref{hom}. The visibility of this fit is $V=0.949\pm0.004$. To correctly determine the degree of quantum interference in our devices we measured the rate of detection of two photons that were created in two separate pairs (blue line). Such events arise due to the relatively long (5~ns) detection window \cite{hifi-note-2}. 
This rate was experimentally determined by measuring two-fold detections with a difference in arrival time of $>$~5~ns, so as to detect only photons generated in separate pairs 
\cite{hifi-note-3}. The green line shows the minimum for perfect quantum interference in an $\eta=0.5267 \pm 0.0004$ coupler, given the measured rate of different pair events. The quantum interference visibility taking this rate into account is $V_{meas}=0.995 \pm 0.004$ which corresponds to  a relative visibility of $V_{rel} \equiv V_{meas}/V_{ideal}= 1.001 \pm 0.004\%$.
This directional coupler therefore shows ideal quantum interference, to within small error bars \cite{hifi-note-4}.

In addition to this high-fidelity quantum interference, general quantum photonic circuits consist of quantum interferometers coupled to classical interferometers operating at the single photon level.  The CNOT gate shown in Fig. \ref{wgs} is therefore an ideal benchmarking device as it contains all the elements of generalized circuits, and its performance therefore shows what can be achieved for such circuits. This gate is designed to work with probability 1/9---the presence of only one photon in the control and one photon in the target signals success of the gate \cite{ra-pra-65-062324,ho-pra-66-024308,ob-nat-426-264,ob-prl-93-080502}. The circuit's performance is quantified \cite{wh-josab-24-172} by the `truth table' shown in Fig.~\ref{truth}(c), taking into account the rate of detecting photons from different pairs, as described above. It has an average correct output probability probability or `logical basis fidelity' of $F=0.969\pm 0.002$ with CNOT operation [Fig. 4(a)].

The measured $\eta$'s of our device differed slightly from the values shown in Fig.~\ref{wgs}: We measured the `1/2' couplers to be $\eta=0.442\pm 0.001$ and $\eta=0.452\pm 0.001$, and the `1/3' coupler in the control part of the circuit to be $\eta=0.3078\pm 0.0009$ \cite{hifi-note-5}; we are not able to directly measure the reflectivities of the two lower `1/3' couplers shown in Fig.~\ref{wgs} because they are embedded in the circuit \cite{hifi-note-6}. Figure~\ref{truth}(b) shows the ideal operation expected for these $\eta$ values, assuming all `1/3' couplers are $\eta=0.3078$.
To quantify the overlap between the ideal $I$ and measured $M$ operation we use the similarity: 
\begin{equation}
\label{compare}
S=\frac{\left(\sum  _{i,j=1}^{4}\sqrt{I_{i,j}M_{i,j}}\right)^2}{16}, 
\end{equation}
which is a generalisation of the average fidelity based on the (classical) fidelity between probability distributions \cite{pr-prl-92-190402,ra-pra-73-012113,cl-pra-79-030303}, and obtain $S=0.993 \pm 0.002$.
If we allow a $\pm1$\% variation of the $\eta$'s for the lower `1/3' couplers, which is a large range given the data, we still obtain $S\ge99$\%; the worst case is 98.9\% \cite{hifi-note-7}.

The results presented here demonstrate that photonic quantum circuits can operate with very high fidelities: worst case operation of the devices described here is in the $10^{-2}-10^{-3}$ range \cite{hifi-note-8}. All linear optical quantum circuits are composed of the quantum and classical interferometers demonstrated here; we can therefore expect the same performance levels from general circuits fabricated in this way. We stress that here we have been concerned with the performance of the photonic quantum circuits themselves, although quantifying this performance required construction of a spectrally tuned SPDC pair photon source.
Requirements for single photon source and detector efficiencies are promising \cite{va-prl-100-060502}. A key challenge for on-demand single photon sources will be to produce photonic qubits with a high degree of indistinguishability, as demonstrated here and verified by quantum interference. Combined with the results presented here, high-efficiency sources and detectors will enable fault tolerant quantum circuit operation across the spectrum of photonic quantum information science and technology applications.
\vspace{6pt}

We thank J.C.F. Matthews for helpful comments. 
This work was supported by EPSRC, ERC, the Leverhulme Trust, QIP IRC, IARPA and NSQI.
J.L.O'B. acknowledges a Royal Society Wolfson Merit Award.

\subsection*{Appendix}

A 402 nm CW laser was focused to a waist of ~40 $\mu$m in a 2 mm thick biaxial Type-I BiBO (BiB$_3$O$_6$) crystal.  The crystal was cut to generate photon pairs at the degenerate wavelength of 804 nm at an opening half angle of 3$^{\circ}$.  In contrast to the Type-II case [Kursiefer et al Phys. Rev. A 64 023802], we assume that the waist of these `daughter' photons is the same as the waist of the pump beam in the crystal, since the spatial-spectral spread is much less for Type-I. The daughter photons with the calculated waist at the crystal were then collected into PMF fibres using a single 11 mm aspheric lens, after passing through interference filters with a 2 nm FWHM, centred at 804 nm.

The two main factors in obtaining indistinguishable photon pairs from this source were ensuring that they share the same spectral mode and the same polarization, both of which should be pure. Since the spectrum of the emitted photons is a function of angular spread and the tilt of the interference filters these should be matched, and this was achieved by measuring the transmitted spectra of the photons. The spatial collection point and the tilt of the interference filter were matched so that the profile of the transmitted photon had a Gaussian shape.  The Gaussian profiles of both photons were centred at the same (degenerate) wavelength.

Purity in the polarisation of the photons was achieved without polarizing optics. Phase matching conditions dictate that the polarization of the daughter photons are orthogonal to the principal plane, and to maximise the SPDC process the polarisation of the pump beam should lie in the principal plane [Dmitriev].  To maximise the polarization purity of the photons collected into PMF fibres, either the slow or fast axis of the fibre must be perfectly parallel (or orthogonal) to the principal plane, since PMFs decohere the polarization of any photon that projects onto both axes. A polarization purity test on the photons that exit the PMF confirmed this alignment. Although the biaxial crystal used here has a more complex geometry than a uniaxial crystal, phase matching was restricted to a direction where one can define a principle plane in the same way as a uniaxial crystal.


\begin{thebibliography}{50}
\expandafter\ifx\csname natexlab\endcsname\relax\def\natexlab#1{#1}\fi
\expandafter\ifx\csname bibnamefont\endcsname\relax
  \def\bibnamefont#1{#1}\fi
\expandafter\ifx\csname bibfnamefont\endcsname\relax
  \def\bibfnamefont#1{#1}\fi
\expandafter\ifx\csname citenamefont\endcsname\relax
  \def\citenamefont#1{#1}\fi
\expandafter\ifx\csname url\endcsname\relax
  \def\url#1{\texttt{#1}}\fi
\expandafter\ifx\csname urlprefix\endcsname\relax\def\urlprefix{URL }\fi
\providecommand{\bibinfo}[2]{#2}
\providecommand{\eprint}[2][]{\url{#2}}

\bibitem[{\citenamefont{Nielsen and Chuang}(2000)}]{nielsen}
\bibinfo{author}{\bibfnamefont{M.~A.} \bibnamefont{Nielsen}} \bibnamefont{and}
  \bibinfo{author}{\bibfnamefont{I.~L.} \bibnamefont{Chuang}},
  \emph{\bibinfo{title}{Quantum Computation and Quantum Information}}
  (\bibinfo{publisher}{Cambridge University Press}, \bibinfo{year}{2000}).

\bibitem[{\citenamefont{Gisin et~al.}(2002)\citenamefont{Gisin, Ribordy,
  Tittel, and Zbinden}}]{gi-rmp-74-145}
\bibinfo{author}{\bibfnamefont{N.}~\bibnamefont{Gisin}},
  \bibinfo{author}{\bibfnamefont{G.}~\bibnamefont{Ribordy}},
  \bibinfo{author}{\bibfnamefont{W.}~\bibnamefont{Tittel}}, \bibnamefont{and}
  \bibinfo{author}{\bibfnamefont{H.}~\bibnamefont{Zbinden}},
  \bibinfo{journal}{Rev. Mod. Phys.} \textbf{\bibinfo{volume}{74}},
  \bibinfo{pages}{145} (\bibinfo{year}{2002}).

\bibitem[{\citenamefont{Gisin and Thew}(2007)}]{gi-nphot-1-165}
\bibinfo{author}{\bibfnamefont{N.}~\bibnamefont{Gisin}} \bibnamefont{and}
  \bibinfo{author}{\bibfnamefont{R.}~\bibnamefont{Thew}},
  \bibinfo{journal}{Nature Photon.} \textbf{\bibinfo{volume}{1}},
  \bibinfo{pages}{165} (\bibinfo{year}{2007}).

\bibitem[{\citenamefont{Deutsch}(1985)}]{de-prsla-400-97}
\bibinfo{author}{\bibfnamefont{D.}~\bibnamefont{Deutsch}},
  \bibinfo{journal}{Proc. R. Soc. Lond. A} \textbf{\bibinfo{volume}{400}},
  \bibinfo{pages}{97} (\bibinfo{year}{1985}).

\bibitem[{\citenamefont{Ladd and et~al.}(2010)}]{la-nat-464-45}
\bibinfo{author}{\bibfnamefont{T.~D.} \bibnamefont{Ladd}} \bibnamefont{and}
  \bibinfo{author}{\bibnamefont{et~al.}}, \bibinfo{journal}{Nature}
  \textbf{\bibinfo{volume}{464}}, \bibinfo{pages}{45} (\bibinfo{year}{2010}).

\bibitem[{\citenamefont{Giovannetti et~al.}(2004)\citenamefont{Giovannetti,
  Lloyd, and Maccone}}]{gi-sci-306-1330}
\bibinfo{author}{\bibfnamefont{V.}~\bibnamefont{Giovannetti}},
  \bibinfo{author}{\bibfnamefont{S.}~\bibnamefont{Lloyd}}, \bibnamefont{and}
  \bibinfo{author}{\bibfnamefont{L.}~\bibnamefont{Maccone}},
  \bibinfo{journal}{Science} \textbf{\bibinfo{volume}{306}},
  \bibinfo{pages}{1330} (\bibinfo{year}{2004}).

\bibitem[{\citenamefont{Knill}(2005)}]{kn-nat-434-39}
\bibinfo{author}{\bibfnamefont{E.}~\bibnamefont{Knill}},
  \bibinfo{journal}{Nature} \textbf{\bibinfo{volume}{434}}, \bibinfo{pages}{39}
  (\bibinfo{year}{2005}).

\bibitem[{\citenamefont{Dawson et~al.}(2006{\natexlab{a}})\citenamefont{Dawson,
  Haselgrove, and Nielsen}}]{da-prl-96-020501}
\bibinfo{author}{\bibfnamefont{C.~M.} \bibnamefont{Dawson}},
  \bibinfo{author}{\bibfnamefont{H.~L.} \bibnamefont{Haselgrove}},
  \bibnamefont{and} \bibinfo{author}{\bibfnamefont{M.~A.}
  \bibnamefont{Nielsen}}, \bibinfo{journal}{Phys. Rev. Lett.}
  \textbf{\bibinfo{volume}{96}}, \bibinfo{eid}{020501}
  (\bibinfo{year}{2006}{\natexlab{a}}).

\bibitem[{\citenamefont{Raussendorf and Harrington}(2007)}]{ra-prl-98-190504}
\bibinfo{author}{\bibfnamefont{R.}~\bibnamefont{Raussendorf}} \bibnamefont{and}
  \bibinfo{author}{\bibfnamefont{J.}~\bibnamefont{Harrington}},
  \bibinfo{journal}{Phys. Rev. Lett.} \textbf{\bibinfo{volume}{98}},
  \bibinfo{eid}{190504} (\bibinfo{year}{2007}).

\bibitem[{\citenamefont{Dawson et~al.}(2006{\natexlab{b}})\citenamefont{Dawson,
  Haselgrove, and Nielsen}}]{da-pra-73-052306}
\bibinfo{author}{\bibfnamefont{C.~M.} \bibnamefont{Dawson}},
  \bibinfo{author}{\bibfnamefont{H.~L.} \bibnamefont{Haselgrove}},
  \bibnamefont{and} \bibinfo{author}{\bibfnamefont{M.~A.}
  \bibnamefont{Nielsen}}, \bibinfo{journal}{Phys. Rev. A}
  \textbf{\bibinfo{volume}{73}}, \bibinfo{eid}{052306}
  (\bibinfo{year}{2006}{\natexlab{b}}).

\bibitem[{\citenamefont{Jan~Benhelm and Blatt}(2007)}]{be-natphys-4-463}
\bibinfo{author}{\bibfnamefont{C.~F.~R.} \bibnamefont{Jan~Benhelm},
  \bibfnamefont{Gerhard~Kirchmair}} \bibnamefont{and}
  \bibinfo{author}{\bibfnamefont{R.}~\bibnamefont{Blatt}},
  \bibinfo{journal}{Nature Phys.} \textbf{\bibinfo{volume}{4}},
  \bibinfo{pages}{463} (\bibinfo{year}{2007}).

\bibitem[{\citenamefont{O'Brien et~al.}(2009)\citenamefont{O'Brien, Furusawa,
  and Vu\v{c}kovi\'{c}}}]{ob-nphot-3-687}
\bibinfo{author}{\bibfnamefont{J.~L.} \bibnamefont{O'Brien}},
  \bibinfo{author}{\bibfnamefont{A.}~\bibnamefont{Furusawa}}, \bibnamefont{and}
  \bibinfo{author}{\bibfnamefont{J.}~\bibnamefont{Vu\v{c}kovi\'{c}}},
  \bibinfo{journal}{Nature Photon.} \textbf{\bibinfo{volume}{3}},
  \bibinfo{pages}{687} (\bibinfo{year}{2009}).

\bibitem[{\citenamefont{Knill et~al.}(2001)\citenamefont{Knill, Laflamme, and
  Milburn}}]{kn-nat-409-46}
\bibinfo{author}{\bibfnamefont{E.}~\bibnamefont{Knill}},
  \bibinfo{author}{\bibfnamefont{R.}~\bibnamefont{Laflamme}}, \bibnamefont{and}
  \bibinfo{author}{\bibfnamefont{G.~J.} \bibnamefont{Milburn}},
  \bibinfo{journal}{Nature} \textbf{\bibinfo{volume}{409}}, \bibinfo{pages}{46}
  (\bibinfo{year}{2001}).

\bibitem[{\citenamefont{O'Brien}(2007)}]{ob-sci-318-1567}
\bibinfo{author}{\bibfnamefont{J.~L.} \bibnamefont{O'Brien}},
  \bibinfo{journal}{Science} \textbf{\bibinfo{volume}{318}},
  \bibinfo{pages}{1567} (\bibinfo{year}{2007}).

\bibitem[{\citenamefont{Politi et~al.}(2008)\citenamefont{Politi, Cryan,
  Rarity, Yu, and O'Brien}}]{po-sci-320-646}
\bibinfo{author}{\bibfnamefont{A.}~\bibnamefont{Politi}},
  \bibinfo{author}{\bibfnamefont{M.~J.} \bibnamefont{Cryan}},
  \bibinfo{author}{\bibfnamefont{J.~G.} \bibnamefont{Rarity}},
  \bibinfo{author}{\bibfnamefont{S.}~\bibnamefont{Yu}}, \bibnamefont{and}
  \bibinfo{author}{\bibfnamefont{J.~L.} \bibnamefont{O'Brien}},
  \bibinfo{journal}{Science} \textbf{\bibinfo{volume}{320}},
  \bibinfo{pages}{646} (\bibinfo{year}{2008}).

\bibitem[{\citenamefont{Politi et~al.}(2009)\citenamefont{Politi, Matthews, and
  O'Brien}}]{po-sci-325-1221}
\bibinfo{author}{\bibfnamefont{A.}~\bibnamefont{Politi}},
  \bibinfo{author}{\bibfnamefont{J.~C.~F.} \bibnamefont{Matthews}},
  \bibnamefont{and} \bibinfo{author}{\bibfnamefont{J.~L.}
  \bibnamefont{O'Brien}}, \bibinfo{journal}{Science}
  \textbf{\bibinfo{volume}{325}}, \bibinfo{pages}{1221} (\bibinfo{year}{2009}).

\bibitem[{\citenamefont{Marshall et~al.}(2009)\citenamefont{Marshall, Politi,
  Matthews, Dekker, Ams, Withford, and O'Brien}}]{ma-oe-17-12546}
\bibinfo{author}{\bibfnamefont{G.~D.} \bibnamefont{Marshall}},
  \bibinfo{author}{\bibfnamefont{A.}~\bibnamefont{Politi}},
  \bibinfo{author}{\bibfnamefont{J.~C.~F.} \bibnamefont{Matthews}},
  \bibinfo{author}{\bibfnamefont{P.}~\bibnamefont{Dekker}},
  \bibinfo{author}{\bibfnamefont{M.}~\bibnamefont{Ams}},
  \bibinfo{author}{\bibfnamefont{M.~J.} \bibnamefont{Withford}},
  \bibnamefont{and} \bibinfo{author}{\bibfnamefont{J.~L.}
  \bibnamefont{O'Brien}}, \bibinfo{journal}{Opt. Express}
  \textbf{\bibinfo{volume}{17}}, \bibinfo{pages}{12546} (\bibinfo{year}{2009}).

\bibitem[{\citenamefont{Matthews et~al.}(2009)\citenamefont{Matthews, Politi,
  Stefanov, and O'Brien}}]{matthews-2008}
\bibinfo{author}{\bibfnamefont{J.~C.~F.} \bibnamefont{Matthews}},
  \bibinfo{author}{\bibfnamefont{A.}~\bibnamefont{Politi}},
  \bibinfo{author}{\bibfnamefont{A.}~\bibnamefont{Stefanov}}, \bibnamefont{and}
  \bibinfo{author}{\bibfnamefont{J.~L.} \bibnamefont{O'Brien}},
  \bibinfo{journal}{Nature Photon.} \textbf{\bibinfo{volume}{3}},
  \bibinfo{pages}{346} (\bibinfo{year}{2009}).

\bibitem[{\citenamefont{Ralph et~al.}(2001)\citenamefont{Ralph, Langford, Bell,
  and White}}]{ra-pra-65-062324}
\bibinfo{author}{\bibfnamefont{T.~C.} \bibnamefont{Ralph}},
  \bibinfo{author}{\bibfnamefont{N.~K.} \bibnamefont{Langford}},
  \bibinfo{author}{\bibfnamefont{T.~B.} \bibnamefont{Bell}}, \bibnamefont{and}
  \bibinfo{author}{\bibfnamefont{A.~G.} \bibnamefont{White}},
  \bibinfo{journal}{Phys. Rev. A} \textbf{\bibinfo{volume}{65}},
  \bibinfo{pages}{062324} (\bibinfo{year}{2001}).

\bibitem[{\citenamefont{Hofmann and Takeuchi}(2001)}]{ho-pra-66-024308}
\bibinfo{author}{\bibfnamefont{H.~F.} \bibnamefont{Hofmann}} \bibnamefont{and}
  \bibinfo{author}{\bibfnamefont{S.}~\bibnamefont{Takeuchi}},
  \bibinfo{journal}{Phys. Rev. A} \textbf{\bibinfo{volume}{66}},
  \bibinfo{pages}{024308} (\bibinfo{year}{2001}).

\bibitem[{\citenamefont{O'Brien et~al.}(2003)\citenamefont{O'Brien, Pryde,
  White, Ralph, and Branning}}]{ob-nat-426-264}
\bibinfo{author}{\bibfnamefont{J.~L.} \bibnamefont{O'Brien}},
  \bibinfo{author}{\bibfnamefont{G.~J.} \bibnamefont{Pryde}},
  \bibinfo{author}{\bibfnamefont{A.~G.} \bibnamefont{White}},
  \bibinfo{author}{\bibfnamefont{T.~C.} \bibnamefont{Ralph}}, \bibnamefont{and}
  \bibinfo{author}{\bibfnamefont{D.}~\bibnamefont{Branning}},
  \bibinfo{journal}{Nature} \textbf{\bibinfo{volume}{426}},
  \bibinfo{pages}{264} (\bibinfo{year}{2003}).

\bibitem[{\citenamefont{O'Brien et~al.}(2004)\citenamefont{O'Brien, Pryde,
  Gilchrist, James, Langford, Ralph, and White}}]{ob-prl-93-080502}
\bibinfo{author}{\bibfnamefont{J.~L.} \bibnamefont{O'Brien}},
  \bibinfo{author}{\bibfnamefont{G.~J.} \bibnamefont{Pryde}},
  \bibinfo{author}{\bibfnamefont{A.}~\bibnamefont{Gilchrist}},
  \bibinfo{author}{\bibfnamefont{D.~F.~V.} \bibnamefont{James}},
  \bibinfo{author}{\bibfnamefont{N.~K.} \bibnamefont{Langford}},
  \bibinfo{author}{\bibfnamefont{T.~C.} \bibnamefont{Ralph}}, \bibnamefont{and}
  \bibinfo{author}{\bibfnamefont{A.~G.} \bibnamefont{White}},
  \bibinfo{journal}{Phys. Rev. Lett.} \textbf{\bibinfo{volume}{93}},
  \bibinfo{eid}{080502} (\bibinfo{year}{2004}).

\bibitem[{\citenamefont{Hong et~al.}(1987)\citenamefont{Hong, Ou, and
  Mandel}}]{ho-prl-59-2044}
\bibinfo{author}{\bibfnamefont{C.~K.} \bibnamefont{Hong}},
  \bibinfo{author}{\bibfnamefont{Z.~Y.} \bibnamefont{Ou}}, \bibnamefont{and}
  \bibinfo{author}{\bibfnamefont{L.}~\bibnamefont{Mandel}},
  \bibinfo{journal}{Phys. Rev. Lett.} \textbf{\bibinfo{volume}{59}},
  \bibinfo{pages}{2044} (\bibinfo{year}{1987}).

\bibitem[{\citenamefont{Shor}(1996)}]{shor96}
\bibinfo{author}{\bibfnamefont{P.}~\bibnamefont{Shor}}, in
  \emph{\bibinfo{booktitle}{Proceedings, {37\textsuperscript{th}} Annual
  Symposium on Fundamentals of Computer Science}} (\bibinfo{publisher}{IEEE
  Press}, \bibinfo{year}{1996}), p.~\bibinfo{pages}{56}.

\bibitem[{\citenamefont{Steane}(1996)}]{st-prl-77-793}
\bibinfo{author}{\bibfnamefont{A.~M.} \bibnamefont{Steane}},
  \bibinfo{journal}{Phys. Rev. Lett.} \textbf{\bibinfo{volume}{77}},
  \bibinfo{pages}{793} (\bibinfo{year}{1996}).

\bibitem[{\citenamefont{Knill et~al.}(1998)\citenamefont{Knill, Laflamme, and
  Zurek}}]{kn-prsa-454-365}
\bibinfo{author}{\bibfnamefont{E.}~\bibnamefont{Knill}},
  \bibinfo{author}{\bibfnamefont{R.}~\bibnamefont{Laflamme}}, \bibnamefont{and}
  \bibinfo{author}{\bibfnamefont{W.~H.} \bibnamefont{Zurek}},
  \bibinfo{journal}{Proc. R. Soc. London A} \textbf{\bibinfo{volume}{454}},
  \bibinfo{pages}{365} (\bibinfo{year}{1998}).

\bibitem[{\citenamefont{Varnava et~al.}(2008)\citenamefont{Varnava, Browne, and
  Rudolph}}]{va-prl-100-060502}
\bibinfo{author}{\bibfnamefont{M.}~\bibnamefont{Varnava}},
  \bibinfo{author}{\bibfnamefont{D.~E.} \bibnamefont{Browne}},
  \bibnamefont{and} \bibinfo{author}{\bibfnamefont{T.}~\bibnamefont{Rudolph}},
  \bibinfo{journal}{Phys. Rev. Lett.} \textbf{\bibinfo{volume}{100}},
  \bibinfo{eid}{060502} (\bibinfo{year}{2008}).

\bibitem[{hif({\natexlab{a}})}]{hifi-note-9}
\bibinfo{note}{See EPAPS for details.}

\bibitem[{hif({\natexlab{b}})}]{hifi-note-1}
\bibinfo{note}{In the case of non-linear `spontaneous parametric
  downconversion' SPDC sources, multiphoton effects are also an important
  source of error \cite{weinhold-2008}.}

\bibitem[{sps()}]{sps}
\bibinfo{note}{Focus on Single Photons on Demand, Eds. P. Grangier, B. Sanders,
  and J. Vuckovic, New J. Phys. 6 (2004)}.

\bibitem[{\citenamefont{Shields}(2007)}]{sh-nphot-1-215}
\bibinfo{author}{\bibfnamefont{A.~J.} \bibnamefont{Shields}},
  \bibinfo{journal}{Nature Photon.} \textbf{\bibinfo{volume}{1}},
  \bibinfo{pages}{215} (\bibinfo{year}{2007}).

\bibitem[{spd()}]{spd}
\bibinfo{note}{Single-photon detectors, applications, and measuremen, Eds. A.
  Migdal and J. Dowling, J. Mod. Opt. 51 (2004)}.

\bibitem[{\citenamefont{Hadfiled}(2009)}]{ha-nphot-3-696}
\bibinfo{author}{\bibfnamefont{R.~H.} \bibnamefont{Hadfiled}},
  \bibinfo{journal}{Nature Photon.} \textbf{\bibinfo{volume}{3}},
  \bibinfo{pages}{696} (\bibinfo{year}{2009}).

\bibitem[{\citenamefont{Smith et~al.}(2009)\citenamefont{Smith, Kundys,
  Thomas-Peter, Smith, and Walmsley}}]{sm-oe-17-13516}
\bibinfo{author}{\bibfnamefont{B.~J.} \bibnamefont{Smith}},
  \bibinfo{author}{\bibfnamefont{D.}~\bibnamefont{Kundys}},
  \bibinfo{author}{\bibfnamefont{N.}~\bibnamefont{Thomas-Peter}},
  \bibinfo{author}{\bibfnamefont{P.~G.~R.} \bibnamefont{Smith}},
  \bibnamefont{and} \bibinfo{author}{\bibfnamefont{I.~A.}
  \bibnamefont{Walmsley}}, \bibinfo{journal}{Opt. Express}
  \textbf{\bibinfo{volume}{17}}, \bibinfo{pages}{13516} (\bibinfo{year}{2009}).

\bibitem[{\citenamefont{Sanaka et~al.}(2006)\citenamefont{Sanaka, Resch, and
  Zeilinger}}]{sa-prl-96-083601}
\bibinfo{author}{\bibfnamefont{K.}~\bibnamefont{Sanaka}},
  \bibinfo{author}{\bibfnamefont{K.~J.} \bibnamefont{Resch}}, \bibnamefont{and}
  \bibinfo{author}{\bibfnamefont{A.}~\bibnamefont{Zeilinger}},
  \bibinfo{journal}{Phys. Rev. Lett.} \textbf{\bibinfo{volume}{96}},
  \bibinfo{eid}{083601} (\bibinfo{year}{2006}).

\bibitem[{\citenamefont{Resch et~al.}(2007)\citenamefont{Resch, O'Brien,
  Weinhold, Sanaka, Lanyon, Langford, and White}}]{re-prl-98-203602}
\bibinfo{author}{\bibfnamefont{K.~J.} \bibnamefont{Resch}},
  \bibinfo{author}{\bibfnamefont{J.~L.} \bibnamefont{O'Brien}},
  \bibinfo{author}{\bibfnamefont{T.~J.} \bibnamefont{Weinhold}},
  \bibinfo{author}{\bibfnamefont{K.}~\bibnamefont{Sanaka}},
  \bibinfo{author}{\bibfnamefont{B.~P.} \bibnamefont{Lanyon}},
  \bibinfo{author}{\bibfnamefont{N.~K.} \bibnamefont{Langford}},
  \bibnamefont{and} \bibinfo{author}{\bibfnamefont{A.~G.} \bibnamefont{White}},
  \bibinfo{journal}{Phys. Rev. Lett.} \textbf{\bibinfo{volume}{98}},
  \bibinfo{eid}{203602} (\bibinfo{year}{2007}).

\bibitem[{\citenamefont{Okamoto et~al.}(2009)\citenamefont{Okamoto, O'Brien,
  Hofmann, Nagata, Sasaki, and Takeuchi}}]{ok-sci-323-483}
\bibinfo{author}{\bibfnamefont{R.}~\bibnamefont{Okamoto}},
  \bibinfo{author}{\bibfnamefont{J.~L.} \bibnamefont{O'Brien}},
  \bibinfo{author}{\bibfnamefont{H.~F.} \bibnamefont{Hofmann}},
  \bibinfo{author}{\bibfnamefont{T.}~\bibnamefont{Nagata}},
  \bibinfo{author}{\bibfnamefont{K.}~\bibnamefont{Sasaki}}, \bibnamefont{and}
  \bibinfo{author}{\bibfnamefont{S.}~\bibnamefont{Takeuchi}},
  \bibinfo{journal}{Science} \textbf{\bibinfo{volume}{323}},
  \bibinfo{pages}{483} (\bibinfo{year}{2009}).

\bibitem[{\citenamefont{Langford et~al.}(2005)\citenamefont{Langford, Weinhold,
  Prevedel, Resch, Gilchrist, O'Brien, Pryde, and White}}]{la-prl-95-210504}
\bibinfo{author}{\bibfnamefont{N.~K.} \bibnamefont{Langford}},
  \bibinfo{author}{\bibfnamefont{T.~J.} \bibnamefont{Weinhold}},
  \bibinfo{author}{\bibfnamefont{R.}~\bibnamefont{Prevedel}},
  \bibinfo{author}{\bibfnamefont{K.~J.} \bibnamefont{Resch}},
  \bibinfo{author}{\bibfnamefont{A.}~\bibnamefont{Gilchrist}},
  \bibinfo{author}{\bibfnamefont{J.~L.} \bibnamefont{O'Brien}},
  \bibinfo{author}{\bibfnamefont{G.~J.} \bibnamefont{Pryde}}, \bibnamefont{and}
  \bibinfo{author}{\bibfnamefont{A.~G.} \bibnamefont{White}},
  \bibinfo{journal}{Phys. Rev. Lett.} \textbf{\bibinfo{volume}{95}},
  \bibinfo{eid}{210504} (\bibinfo{year}{2005}).

\bibitem[{hif({\natexlab{c}})}]{hifi-note-2}
\bibinfo{note}{Using a smaller timing window, possible with detectors with a
  small timing jitter \cite{ha-nphot-3-696}, reduces this rate.}

\bibitem[{hif({\natexlab{d}})}]{hifi-note-3}
\bibinfo{note}{We measured this rate at the centre of the dip, which gives a
  slightly lower rate than at other delay times, and is therefore a worst case
  assumption.}

\bibitem[{hif({\natexlab{e}})}]{hifi-note-4}
\bibinfo{note}{Note that no correction for multi-pair emission within a
  coherence time was made; such a correction can be problematic, but is not
  required for a CW laser, since there is a negligible probability to create
  more than one pair of photons within the coherence time of $\sim 10^{-12}$
  s.}

\bibitem[{\citenamefont{White et~al.}(2007)\citenamefont{White, Gilchrist,
  Pryde, O'Brien, Bremner, and Langford}}]{wh-josab-24-172}
\bibinfo{author}{\bibfnamefont{A.~G.} \bibnamefont{White}},
  \bibinfo{author}{\bibfnamefont{A.}~\bibnamefont{Gilchrist}},
  \bibinfo{author}{\bibfnamefont{G.~J.} \bibnamefont{Pryde}},
  \bibinfo{author}{\bibfnamefont{J.~L.} \bibnamefont{O'Brien}},
  \bibinfo{author}{\bibfnamefont{M.~J.} \bibnamefont{Bremner}},
  \bibnamefont{and} \bibinfo{author}{\bibfnamefont{N.~K.}
  \bibnamefont{Langford}}, \bibinfo{journal}{Journal of the Optical Society of
  America B: Optical Physics} \textbf{\bibinfo{volume}{24}},
  \bibinfo{pages}{172} (\bibinfo{year}{2007}).

\bibitem[{hif({\natexlab{f}})}]{hifi-note-5}
\bibinfo{note}{We measured reflectivities by assuming $2\times2$ couplers with
  losses on inputs and outputs; with the constraints of probabilities, we can
  cancel out the effect of losses to obtain the reflectivity.}

\bibitem[{hif({\natexlab{g}})}]{hifi-note-6}
\bibinfo{note}{Note that variable reflectivity beamsplitters have been
  demonstrated in this architecture and used for quantum interference
  \cite{matthews-2008} and could be used to obtain near-to-exact values of
  reflectivity required for any circuit.}

\bibitem[{\citenamefont{Pryde et~al.}(2004)\citenamefont{Pryde, O'Brien, White,
  Bartlett, and Ralph}}]{pr-prl-92-190402}
\bibinfo{author}{\bibfnamefont{G.~J.} \bibnamefont{Pryde}},
  \bibinfo{author}{\bibfnamefont{J.~L.} \bibnamefont{O'Brien}},
  \bibinfo{author}{\bibfnamefont{A.~G.} \bibnamefont{White}},
  \bibinfo{author}{\bibfnamefont{S.~D.} \bibnamefont{Bartlett}},
  \bibnamefont{and} \bibinfo{author}{\bibfnamefont{T.~C.} \bibnamefont{Ralph}},
  \bibinfo{journal}{Phys. Rev. Lett.} \textbf{\bibinfo{volume}{92}},
  \bibinfo{pages}{190402} (\bibinfo{year}{2004}).

\bibitem[{\citenamefont{Ralph et~al.}(2006)\citenamefont{Ralph, Bartlett,
  O'Brien, Pryde, and Wiseman}}]{ra-pra-73-012113}
\bibinfo{author}{\bibfnamefont{T.~C.} \bibnamefont{Ralph}},
  \bibinfo{author}{\bibfnamefont{S.~D.} \bibnamefont{Bartlett}},
  \bibinfo{author}{\bibfnamefont{J.~L.} \bibnamefont{O'Brien}},
  \bibinfo{author}{\bibfnamefont{G.~J.} \bibnamefont{Pryde}}, \bibnamefont{and}
  \bibinfo{author}{\bibfnamefont{H.~M.} \bibnamefont{Wiseman}},
  \bibinfo{journal}{Phys. Rev. A} \textbf{\bibinfo{volume}{73}},
  \bibinfo{eid}{012113} (\bibinfo{year}{2006}).

\bibitem[{\citenamefont{Clark et~al.}(2009)\citenamefont{Clark, Fulconis,
  Rarity, Wadsworth, and O'Brien}}]{cl-pra-79-030303}
\bibinfo{author}{\bibfnamefont{A.~S.} \bibnamefont{Clark}},
  \bibinfo{author}{\bibfnamefont{J.}~\bibnamefont{Fulconis}},
  \bibinfo{author}{\bibfnamefont{J.~G.} \bibnamefont{Rarity}},
  \bibinfo{author}{\bibfnamefont{W.~J.} \bibnamefont{Wadsworth}},
  \bibnamefont{and} \bibinfo{author}{\bibfnamefont{J.~L.}
  \bibnamefont{O'Brien}}, \bibinfo{journal}{Phys. Rev. A}
  \textbf{\bibinfo{volume}{79}}, \bibinfo{eid}{030303} (\bibinfo{year}{2009}).

\bibitem[{hif({\natexlab{h}})}]{hifi-note-7}
\bibinfo{note}{Fig.~\ref{truth} does not provide full characterization of the
  ÔunknownÕ two-qubit quantum process, composed of unitaries, projective
  measurements and decohering processes. However, the data shown here and in
  previous works \cite{po-sci-320-646,matthews-2008} demonstrates that these
  devices preserve coherence. Assuming the device preserves purity restricts it
  to a class of unitaries where the only unknowns are the (trivial) phases
  between computational basis states in the input and output of the device.}

\bibitem[{hif({\natexlab{i}})}]{hifi-note-8}
\bibinfo{note}{Exact conversion between interference visibility or similarity
  and EPGs are beyond the scope of this work, but has previously been
  considered \cite{weinhold-2008}.}

\bibitem[{\citenamefont{Weinhold et~al.}(2008)\citenamefont{Weinhold,
  Gilchrist, Resch, Doherty, O'Brien, Pryde, and White}}]{weinhold-2008}
\bibinfo{author}{\bibfnamefont{T.~J.} \bibnamefont{Weinhold}},
  \bibinfo{author}{\bibfnamefont{A.}~\bibnamefont{Gilchrist}},
  \bibinfo{author}{\bibfnamefont{K.~J.} \bibnamefont{Resch}},
  \bibinfo{author}{\bibfnamefont{A.~C.} \bibnamefont{Doherty}},
  \bibinfo{author}{\bibfnamefont{J.~L.} \bibnamefont{O'Brien}},
  \bibinfo{author}{\bibfnamefont{G.~J.} \bibnamefont{Pryde}}, \bibnamefont{and}
  \bibinfo{author}{\bibfnamefont{A.~G.} \bibnamefont{White}},
  \bibinfo{journal}{arXiv.org:0808.0794}  (\bibinfo{year}{2008}).

\end{thebibliography}
\end{document}